# High resolution electron microscopy for heterogeneous catalysis research[†]

Yong Zhu(朱勇)[1, #], Mingquan Xu(许名权)[1, #], Wu Zhou(周武)[1, *]

1. School of Physical Sciences and CAS Center for Excellence in Topological Quantum Computation, University of Chinese Academy of Sciences, Beijing 100049, China

\# These authors contributed equally to this work.

\* E-mail: wuzhou@ucas.ac.cn

† Dedicated to the 80[th] birthday of Prof. Jing Zhu.

**Abstract**

　　Heterogeneous catalysts are the most important catalysts in industrial reactions. Nanocatalysts, with size ranging from hundreds of nanometers to the atomic scale, possess activities that are closely connected to their structural characteristics such as particle size, surface morphology, and three-dimensional topography. Recently, the development of advanced analytical transmission electron microscopy (TEM) techniques, especially quantitative high-angle annular dark-field (HAADF) imaging and high-energy resolution spectroscopy analysis in scanning transmission electron microscopy (STEM) at the atomic scale, strengthens the power of (S)TEM in analyzing the structural/chemical information of heterogeneous catalysts. Three-dimensional reconstruction from two-dimensional projected images and the real-time recording of structural evolution during catalytic reactions using *in-situ* (S)TEM methods further broaden the scope of (S)TEM observation. The atomic-scale structural information obtained from high resolution (S)TEM has proven to be of significance for better understanding and designing of new catalysts with enhanced performance.

**Introduction**

Most industrial catalytic processes occur in heterogeneous systems, the chemical reactions initiate at the interface between the catalyst nanoparticles and the reaction media, whereas mass transport may occur over several atomic layers localized on the exposed surface of the catalysts. Subtle changes of the surface structures at the nanometer or atomic scale would thus have significant effects on the catalytic efficiency. In addition, for supported heterogeneous catalysts the interaction between the active metal nanoparticles and the support materials also play a critical role in defining the catalytic performance.[1, 2] Such metal-support interaction has also been used to tune the structure of the supported metal species, with the ultimate goal of constructing "single-atom catalysts" where every single precious metal atom can be dispersed onto the surface of the support materials and participate in catalytic reactions.[3] This has become an important trend in advanced catalyst research, promising a way to increase the utilization efficiency of the noble metal. Meanwhile, this poses a major challenge for unveiling the detailed structure of advanced heterogeneous catalysts, and characterization techniques with atomic resolution and single atom sensitivity become indispensable.

Conventional imaging and spectroscopy by optical or X-ray based techniques only gather the averaged structural and chemical information from a relatively bulk volume and cannot provide local structural/compositional information at the atomic scale. By using high-energy electron beams as the "light source", transmission electron microscopes in principle offer the opportunity to achieve picometer-scale spatial resolution, although in practice the spatial resolution is severely limited by many factors including aberrations from the imperfect magnetic lens.[4] With the help of advanced aberration correctors, the state-of-the-art (scanning) transmission electron microscopes ((S)TEM) can nowadays routinely perform imaging with sub-ångström spatial resolution,[5] and imaging and spectroscopy at the single atom scale have become feasible with aberration-corrected STEM even under low accelerating voltage.[6, 7] These technical advances enable analytical TEM as a powerful tool for structural analysis of heterogeneous catalysts.

**Part 1. Unveiling the structure of heterogeneous catalysts at the atomic scale**

Heterogeneous catalytic reactions occur primarily on the surface and interface. Thus, as the size of the nanoparticles dispersed on support materials decreases, the increasing exposed surface could provide more active sites for catalytic reactions. The ultimate condition is that individual isolated atoms are dispersed or anchored on the supports, namely, single atom catalysts (SACs) are formed.[3, 8, 9] Abbet et al.[10] studied $Pd_n$ clusters ($1 \leq n \leq 30$) supported on MgO films for the cyclotrimerization of acetylene, and advocated that "One atom is enough!". Qiao et al.[11] reported a single-atom catalyst for CO oxidation and revealed isolated Pt atoms dispersed on a $FeO_x$ support by high-angle annular dark-field (HAADF) imaging in STEM. Dispersing noble metals atomically on the support materials can also dramatically reduce the amount of costly materials being used, bringing economic benefits.

The emergence of SACs brings forward new challenges in understanding the structure-property relationships for heterogeneous catalysts, because the signals from single atoms are too weak to be detected by conventional characterization techniques. Hence, development of new imaging methods to directly probe the precise location of individual atoms with ultrahigh spatial resolution and spectroscopy methods to collect chemical information with single-atom sensitivity is of particular importance. STEM-HAADF imaging has an intrinsically higher resolution than coherent HRTEM and can directly represent heavy atoms on relatively light supports, due to its incoherent Z-contrast nature. Advanced STEM-HAADF imaging has already achieved a sub-ångström resolution with the adoption of aberration correctors, which has been widely applied to characterizing SAC systems, such as individual Pd atoms on ultrathin $TiO_2$ sheets,[12] isolated Ni sites on MOFs[13] and single Fe dispersed in Fe-N-C catalyst.[14]

By correcting the geometry aberrations of the probe forming lenses, aberration correctors in STEM decrease the probe size and help to achieve a high probe current in a sub-ångström probe, which provides possibility to identify single Si and Pt atoms on graphene via energy dispersive X-ray spectroscopy (EDXS).[7] In addition, electron energy loss spectroscopy (EELS) in STEM has become a powerful method for atomic-scale chemical identification and measurement of bonding information, such as determination of two different bonding configurations from individual impurity Si atoms in graphene,[6] confirming single nitrogen dopants in graphene[15] and mapping out the spatial distributions of individual Li atoms in carbon nanotubes.[16] Here we will discuss a few recent examples to illustrate the application of STEM imaging and spectroscopy analysis on the studies of single atom catalysts.

Low-temperature hydrogen production is important for polymer electrolyte membrane fuel cells (PEMFCs) which are considered as a kind of zero emission devices. In order to accelerate the rate of $H_2$ production, water and carbon-containing sources should both be activated efficiently, and supported catalysts with bifunctional structures have attracted intense research interests in this field. Recently, Ding Ma and co-workers reported two interesting studies on noble metal catalysts supported on α-MoC for low-temperature hydrogen production, where the atomic structures of the active metal species are directly revealed using atomic-resolution STEM-HAADF imaging. In the first study, Lin et al. synthesized a series of platinum (Pt)/α-MoC catalysts as well as Pt on different supports, and found out that the Pt/α-MoC catalyst with 0.2% Pt loading showed the best $H_2$-producing catalytic activity at low temperature, owing to its single Pt atom dispersion on α-MoC support.[17] Atomic-resolution STEM-HAADF images of the 0.2% Pt/α-MoC catalyst verified the fact that Pt metal disperses atomically on the α-MoC surface, as highlighted in red in Fig. 1(a). On the spent catalyst, the atomic dispersion of Pt was well retained (Fig. 1(b)), attributed to the strong interaction between platinum atoms and the α-MoC support. Density functional theory (DFT) calculations revealed that this strong interplay facilitates the formation of electron-rich regions at the Pt-MoC interface which contribute to the high catalytic activity for the hydrogen production reactions.

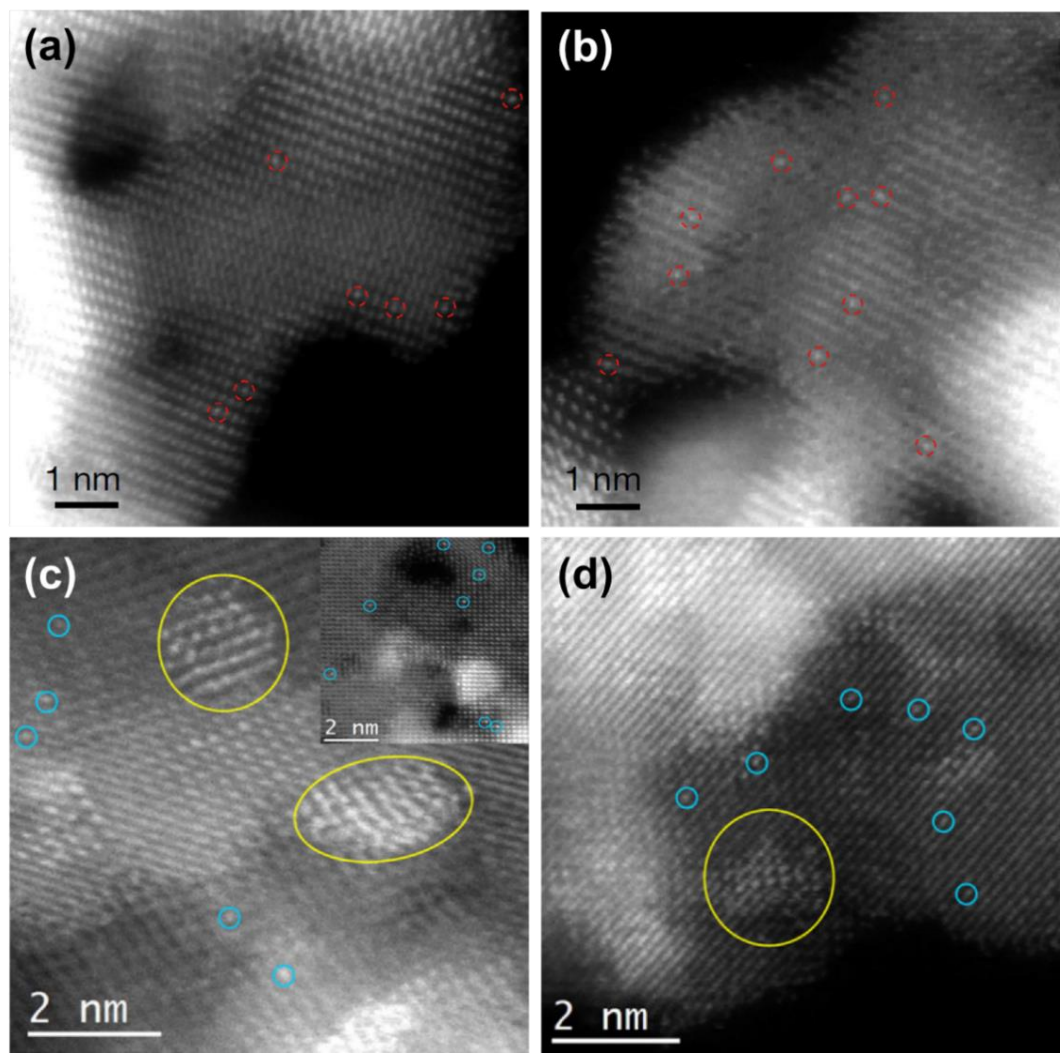

**Fig. 1. STEM-HAADF characterization of highly dispersed noble metals on MoC support.** High-resolution STEM-HAADF images of fresh 0.2% Pt/α-MoC (a) and used 0.2% Pt/α-MoC catalysts (b).[17] High-resolution STEM-HAADF images of fresh 2% Au/α-MoC (c), used 2% Au/α-MoC catalysts (d), and the NaCN-leached specimen (inset in Fig. 1(c)).[18]

The water-gas shift (WGS) reaction is another approach to produce hydrogen at low temperature, and Yao et al. developed an excellent Au/α-MoC catalyst for this reaction.[18] STEM-HAADF imaging was again the key for revealing the atomic structure of the active species in this novel catalyst. STEM-HAADF images show that there are two different Au configurations on the α-MoC surface: individually dispersed Au atoms and Au layered clusters (labeled by blue and yellow in Fig. 1(c), respectively). After catalytic testing, both configurations maintained (Fig. 1(d)), which contributed to the good stability during catalytic process. In order to distinguish the contribution from these two kinds of Au structures, the Au/α-MoC samples were leached by NaCN solution and the corresponding STEM-HAADF image (inset in Fig. 1(c)) showed that Au predominantly existed in an isolated-atom form in the leached sample. The Au-normalized WGS activities of NaCN-leached catalyst dropped dramatically, indicating that the Au clusters, instead of single Au atoms, on α-MoC support provided the major contribution to the low-temperature WGS activity, which is further confirmed by first-principles calculations.

As can be seen from the two examples discussed above, single atoms dispersed on a suitable support surface can indeed form an active catalyst. However, it is not always the case that single atoms with the highest dispersion are catalytically more active than other structures, as illustrated in the case of Au/α-MoC for low temperature WGS. Nevertheless, characterization techniques that can clearly visualize surface structures down to the single atom level is crucial in order to uncover all the possible active site structures and identify the catalytically most active ones. STEM-HAADF imaging provides a feasible way for this purpose, and the atomic structural information obtained from STEM imaging can serve as valuable inputs for first-principles calculations to deduce the correlation between the atomic configuration and catalytic performance, as also illustrated in the two studies discussed above.

Apart from the supported noble metal nanoparticle catalysts, two-dimensional based catalysts are another family of catalysts that require microscopy imaging and spectroscopy analysis with single-atom resolution and sensitivity.

$MoS_2$-based nanomaterials are widely used for industrial hydrodesulfurization reactions, and their catalytic activity is mostly contributed by the edges of the layered materials. It has been well demonstrated that the catalytic activity of $MoS_2$-based catalysts can be promoted by doping the edge sites with other elements (e.g. Co or Ni).[19-21] Identifying the precise location of these substituting dopant atoms can provide us guidance for understanding the catalytic mechanism and designing new catalysts with superb performance. However, conventional (S)TEM techniques with the accelerating voltage exceeding 100 kV generally cause undesired damage to such two-dimensional materials, which may mislead our interpretation of the (S)TEM images. Thus, the application of low-voltage aberration corrected STEM method is crucial for atom-level analysis of beam-sensitive catalysts.

An elegant application of the low-voltage STEM imaging and spectroscopy analysis was reported by Zhu et al., where 60 kV STEM-HAADF imaging, assisted by simultaneous EEL spectrum imaging, was applied to uncover the preferential incorporation of Co-promoter atoms along different crystallographic edges in a Co-promoted $MoS_2$ catalyst.[22] To determine the precise atomic structure and stoichiometry of Co-substituted S edges, sub-Å STEM-HAADF imaging and spatial-resolved EEL spectra were conducted (Fig. 2). By comparing the Mo and Co EELS mapping (Fig. 2(b)) with high-resolution Z-contrast HAADF image (Fig. 2(a)), it could be proved conclusively that the Co atoms substituted Mo-sites at the S-zigzag edges, forming 2S-1Co dumbbells. Furthermore, the S signal became weaker at the outmost edge in the EELS mapping (Fig. 2(c)), and the sulfur $L_{2,3}$ intensity from these outmost S columns exhibited half intensity of that from the inner S columns (Fig. 2(d)), indicating the presence of a single-S-atom row terminated at the S-zigzag edge. Consequently, the Co atoms are tetrahedrally coordinated to S atoms at these decorated edges as illustrated in Fig. 2(e). This intrinsically undercoordinated 1S-1Co edge configuration has been proposed to be active for the adsorption of S-containing reactants. It was also reported that additional Fe $L_{2,3}$ signals were sometimes detected concomitantly with Co signals as shown in Fig. 2(f)-2(h), presumably coming from Fe residues in the graphite support. Due to the neighboring nature of Co and Fe in periodic table, Fe residue may compete with Co for the S edges. This may explain why the catalytic activity of Co-Mo-S catalysts usually show a dependence on the type of carbon supports and demonstrates the importance of the purity of raw materials when synthesizing Co-Mo-S catalysts.

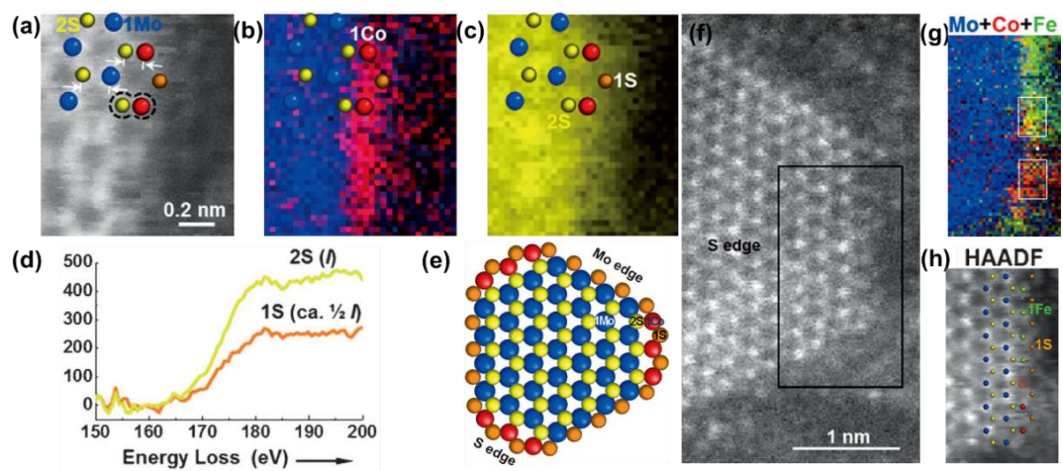

**Fig. 2. STEM imaging and spectroscopy analysis of MoS$_2$ based catalysts at the single atom level.** (a) High-resolution STEM-HAADF image of the S-zigzag edge in Co-Mo-S catalyst and the corresponding EELS mapping (b) of Mo (in blue) and Co (in red) and (c) S (in yellow). (d) The L$_{2,3}$ edges of a 2S and a 1S atomic columns in (c). (e) Schematic of the Co-Mo-S crystal structure. (g) Combined Mo/Co/Fe mapping from the area framed in black in the STEM-HAADF image (f) and (h) corresponding superimposed ball model.[22]

Generally, for bulk MoS$_2$, coordinately saturated sites on the basal planes do not exhibit catalytic activity. However, as the number of layers of MoS$_2$ reduces, the exposed basal plane atoms increase and can become active upon doping. Liu et al. reported a new catalyst by incorporating single Co atoms onto the basal plane of monolayer MoS$_2$ and distinguished Co atom substitution behavior before and after catalytic hydrodeoxygenation (HDO) reaction.[23] STEM-HAADF imaging revealed that some sites in monolayer Co-$^S$MoS$_2$ produce higher image contrast than the nearby Mo and S$_2$ sites as shown in Fig. 3(a), the brighter point was confirmed to be Co atom by HAADF intensity profile (Fig. 3(b)) and EELS line scanning (Fig. 3(c)). Comparing the simulated HAADF image (Fig. 3(g)) of a DFT-optimized structure with the experimental results in Fig. 3(d), it is demonstrated that the Co atoms occupy the Mo-atop site in the fresh Co-$^S$MoS$_2$ catalyst. For the used Co-$^S$MoS$_2$ sample presented in Figs. 3(e) and 3(f), different from the initial Co occupation at the Mo-atop site, another two kinds of substitution sites were detected in HAADF images: Co substituting the sulfur sites (Fig. 3(e)) and Co incorporating in the hollow site in the center of the $MoS_2$ hexagon (Fig. 3(f)). The corresponding simulated images (Figs. 3(h) and 3(i)) from DFT structural optimization confirmed these two atomic configurations. Because there was no loss of Co loading during the catalytic reaction according to inductively coupled plasma (ICP) analysis, this manifests that Co promotor atoms migrate from the original Mo-atop sites to S-substituting sites or hollow sites, creating sulfur vacancies in a close proximity during the catalytic process. The availability of basal active sites due to Co-substitution-induced sulfur vacancies was considered to contribute to the drastic promotion in Co-$^S$MoS$_2$, which provides new insights for designing novel catalysts with high activity and durability.

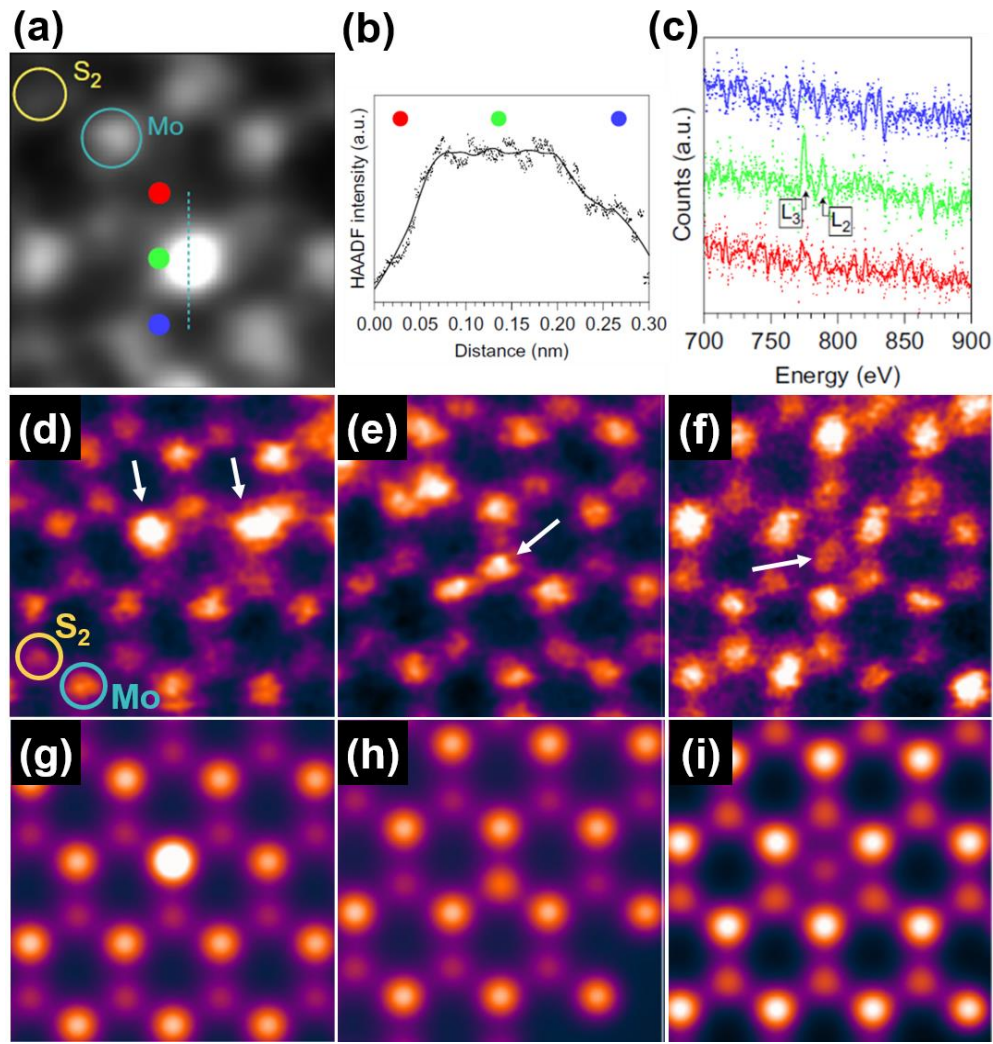

**Fig. 3. STEM-HAADF images and spectra of the Co-$^S$MoS$_2$ catalyst.** (a) The magnified STEM-HAADF image of fresh Co-$^S$MoS$_2$. (b) The corresponding HAADF intensity line scanning in (a). (c) The simultaneous EELS acquired along the line in (a). (d) The HAADF image of Co sitting on the Mo-atop site (marked in white arrow) in fresh catalysts. (e) HAADF image of Co-substituted S site (marked in white arrow) in used catalysts. (f) The HAADF image of Co occupying in hollow site (marked in white arrow) in used catalysts. (g-i) The corresponding simulated HAADF images, respectively.[23]

Oxygen reduction reaction (ORR) plays a key role in polymer electrolyte fuel cells (PEFCs), and noble platinum-group metals have been widely used in this field due to their good oxygen reduction activities, which causes PEFCs costly. Thus, searching for non-precious metal catalysts is significant for large-scale commercialization of fuel cells-powered vehicles. Since Jasinski[24] observed the catalytic activity of cobalt phthalocyanine for ORR in 1964, inexpensive metal-nitrogen-carbon (M-N-C, M=Fe, Co, etc.) compounds[25-27] have been considered as a promising alternative to Pt-based catalysts. In Fe (or Co)-N-C systems, determining the local coordination between metal and nitrogen atoms can provide deeper understanding for their catalytic mechanism, and low-voltage STEM imaging and EELS analysis show advantage in such characterization.

Li et al. synthesized a novel ORR electrocatalyst of few-walled carbon nanotube-graphene (NT-G) complexes and found out that iron impurities and nitrogen doping of this structure are both important to the improved ORR electrocatalytic activity.[28] The STEM-ADF image (Fig. 4(a)) indicated that iron atoms (the highest contrast sites in Fig. 4(b)) are preferentially located on the edge of graphene structure exfoliated from the outer wall of the carbon nanotubes (CNTs). This was further confirmed by the iron EELS mapping in Fig. 4(c). Chemical mapping in Figs. 4(c), (d) and (e) shows that iron and nitrogen atoms on graphene edge are often adjacent to each other, demonstrating possible chemical bonding between Fe and N. Comparing with the results that other

NT-G catalysts without the functionality of Fe (or N) species all exhibited lower ORR activities, it was concluded that Fe-N complex in the NT-G structure act as the active sites during the ORR process.

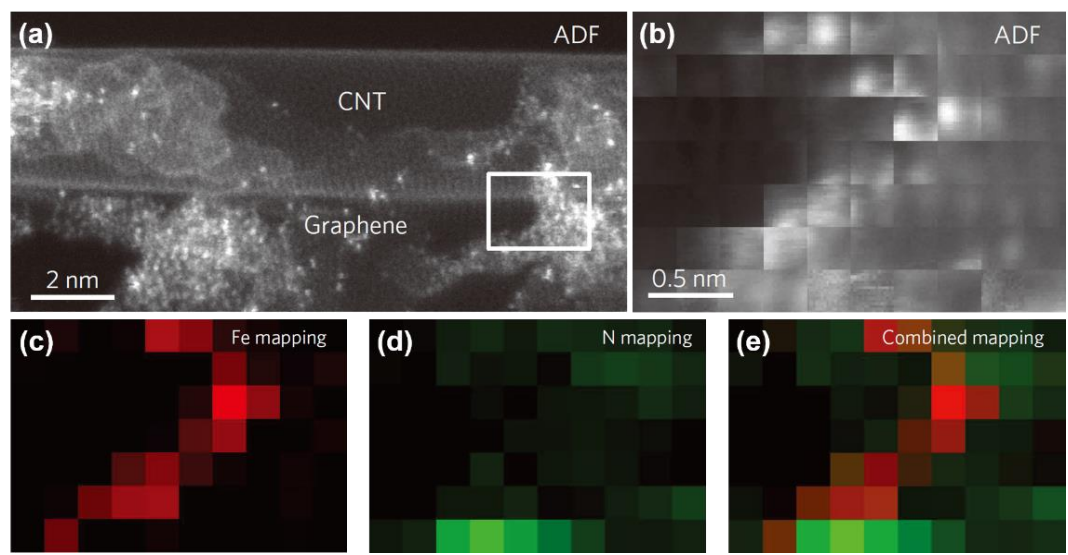

**Fig. 4. STEM imaging and spectrum imaging of a CNT-graphene based catalyst for ORR.** (a) STEM-ADF image of the NT-G complexes. (b) Simultaneously acquired ADF image, (c) Fe EELS mapping, d) N EELS mapping, and (e) the overlaid Fe and N EELS signals from the white framed area in (a).[28]

Limited by the complex NT-G structure and the under-sampling setting of the STEM-EEL spectrum imaging experiment of that time, the abovementioned results did not provide direct visualization of the chemical bonding configuration of the single Fe and N atoms. A more recent study by Chung et al. has moved the study one step forward and proposed FeN$_4$ in graphene lattice as the active site using a combination of STEM-ADF imaging, EELS analysis and DFT calculations.[29] In this work, the (cyanamide+polyaniline)-iron-carbon catalyst ((CM+PANI)-Fe-C)) with a hierarchical porous structure was synthesized, and the atomic resolution STEM-ADF image (Fig. 5(a)) showed that individual Fe atoms are dispersed in the few-layered graphene lattice. EELS spectra acquired from such Fe sites show that N signals are always detected at the Fe atomic sites but are absent in the graphene-only regions (Fig. 5(b)). High resolution EELS elemental mapping also shows the spatial overlapping of Fe and N signals (Figs. 5(d) and 5(e)). The authors claimed that quantification of the EELS spectra indicates an average composition of $FeN_4$ for such sites, although judging from the signal-to-noise ratio of the published spectra we think the error of such quantification could be quite large. Moreover, the observation of STEM imaging revealed that most of the highly dispersed Fe atoms tend to occupy exposed-plane edges and steps of the carbon phases. Though the specific N coordination of these edge-positioned Fe atoms could not be directly discerned with STEM-EELS due to their instability even under the 60 keV electron beam, the authors proposed the higher ORR activities can be expected from edge-hosted $FeN_x$ sites than that of bulk-hosted ones based on quantum chemical calculations.

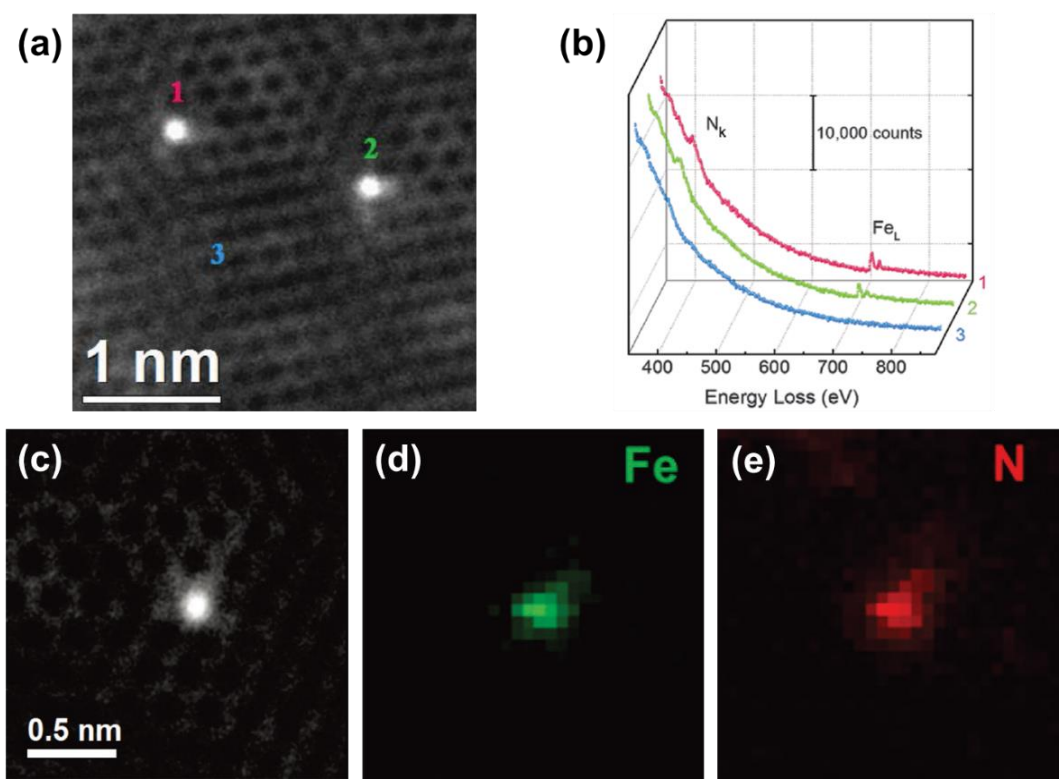

**Fig. 5. Atomic-scale STEM analysis of graphene-based catalysts for ORR.** (a) STEM-HAADF image of individual Fe atoms in few-layered graphene. (b) The EEL spectra of N K-edge and Fe L-edge from the three points indicated in (a). (c) Simultaneously acquired STEM-HAADF image and EELS mapping of Fe (d) and N (e).[29]

In summary, atomic-scale STEM imaging and EELS analysis can efficiently visualize and identify individual heavy atoms in SACs and catalysts based on two-dimensional materials, which can provide direct evidence for uncovering local atomic configurations and the corresponding chemical bonding information. These are important structural information that can be used directly as input models for theoretical calculations for understanding the underlying catalytic mechanism and can ultimately help to design novel catalysts with excellent performance. Readers interested in this topic are encouraged to refer to recent review articles on (S)TEM study of single-atom catalysts.[9, 30]

**Part 2. 3D reconstruction for the study of heterogeneous catalysts**

Images acquired in (S)TEM are two-dimensional projections of the three-dimensional (3D) objects, which inevitably loss structural information in the third dimension.[31, 32] As 3D topography information is important for understanding the exposed facets of catalyst particles and mass transport during catalytic reaction, it is of great importance to develop techniques that can provide structural information in three dimensions, especially for materials that possess complicated morphology and spatial variation of chemical composition.[33]

In 1917 Radon showed mathematically how a series of 2D projection images could be converted back to the 3D structural model through the so-called Radon transformation,[34] and many methods have been developed for 3D reconstruction under TEM and STEM imaging modes to date. Compared with TEM mode, STEM-HAADF tomography is more suitable for crystalline materials by suppressing the diffraction effects and can provide chemical information in 3D reconstruction due to its nature of Z-contrast.[35] There are two main 3D reconstruction approaches under STEM-HAADF imaging mode. The first one, known as depth sectioning method, utilizes a focused electron probe with large convergence angle and a very small depth of focus ($\Delta z$) to acquire a series of through-focal 2D images along the $z$ direction. However, the severe elongation artifact along the $z$-direction due to the missing-cone problem[36] may lead to misinterpretation of the experimental data. The second type is electron tilt tomography, in which a series of 2D projection images of the sample are acquired under different tilt angles and the 3D structure can be reconstructed via different algorithms.

The fidelity of tilt tomography reconstruction method mainly depends on the algorithms being used, the tilt angle range and the number of sampled tilt angles. In many cases, the limited tilt angle due to geometric constraints leads to the "missing wedge" problem and an accompanying decline of resolution in the z-direction,[37, 38] and the electron beam sensitivity of the sample limits the maximum tolerable dose and the number of tilt-series images.[39] These are the major technical challenges faced by electron tilt tomography. Nevertheless, it is still the most commonly used technique for 3D reconstruction in material research nowadays, such as identifying the spatial distribution of second phase/component within the matrix,[40] reconstructing the 3D structure of core/shell nanoparticle[41] and so on.

One important application of 3D reconstruction is for the study of nanoporous catalysts where the inner surface area of interconnected pores can provide more active sites and increase the mass-based activity.[42, 43] Geboes et al. used the STEM-HAADF electron tomography to unveil the 3D porous structure of Pt nanocatalysts synthesized by electrodeposition at different overpotentials.[44] The 3D reconstruction image of the high-overpotential electrodeposited NP in Fig. 6(a) reveals a dendritic surface which seems to have an increased surface-volume ratio, while the corresponding slice (Fig. 6(b)) shows that the small channels in the nanoparticle rarely connect to the outer surface. In contrast, the highly dendritic morphology is present both on the inner and outer surfaces of the low-overpotential synthesized Pt nanoparticle (Figs. 6(c) and 6(d)), which drastically increases the accessible active sites inside the nanopores and enhances the electrochemical surface area normalized ORR activity. The drastic difference of the inner-porous structure in these two kinds of Pt NPs, as revealed by 3D tomography reconstruction, provides the key to unveil the morphology-property relationship.

In another example Xin et al. applied STEM-HAADF tomography to study the porous structure of binary $CoO_x/SiO_2$ nanoparticles, where the HAADF intensity was used to differentiate the heavier $CoO_x$ component from the lighter $SiO_2$.[45] The 3D reconstruction results clearly unveil that the nanocatalyst consists of a $CoO_x$ core with an interconnected nanoporous network and a $SiO_2$ shell, which facilitates the infiltration of gas molecules like hydrogen during catalytic reactions, shown in Fig. 6(e). Coupled with *in-situ* TEM analysis, it was found that the SiO$_2$ coating can mitigate the porosity collapse during the high-temperature reduction process and protect the reduced Co metal core from oxidation. The results can provide a guideline for designing porous supported nanocatalysts.

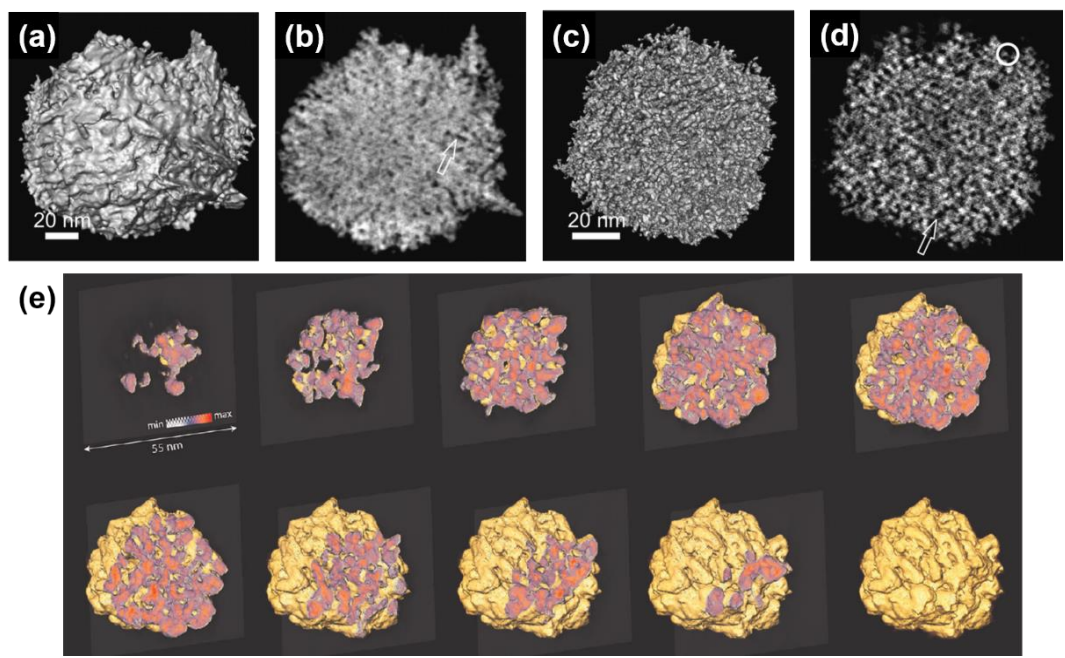

**Fig. 6. STEM-HAADF tomography reconstruction and cross-sections of nanoparticle catalysts.** (a) 3D visualization of the high-overpotential synthesized Pt NPs and (b) a corresponding slice image, (c) 3D visualization of the low-overpotential synthesized Pt NPs and (d) a corresponding slice image.[44] (e) A series of progressing cross-section images of a CoO$_x$/SiO$_2$ nanocatalyst.[45]

The two examples mentioned above used a reconstruction algorithm called simultaneous iterative reconstruction technique (SIRT). The reconstruction quality is mainly determined by the quantity of the 2D projections, and a large number of projections is needed in order to assure a high fidelity.[46] The spatial resolution achievable in these two examples is around 1 nm, which limits more detailed information to be extracted from the 3D reconstruction. Combing discrete tomography, which uses a small number of atomic resolution projections taken along different zone axes of crystalline samples, with new reconstruction algorithms, 3D reconstruction with atomic resolution can be achieved.[47]

Recently, Goris et al. put forward a compressive sensing based reconstruction algorithm, also known as total variation minimization (TVM), to characterize the surface facets of Au nanorods.[48] The TVM method can reconstruct the atomic lattice faithfully from a limited number of projection. Fig. 7(a) shows the reconstructed 3D structure of an Au nanorod from 2D STEM images collected along a few major zone axes ([100], [110], [010] and [1-10]). Figure 7(b) presents the two orthogonal slices extracted from Fig. 7(a). An atomic surface step with a thickness of two atomic layers can be clearly observed at the {001} facet as shown in Fig. 7(c). Furthermore, by applying geometrical phase analysis (GPA) to the atomic-resolution 3D reconstruction, the full 3D strain field can be analyzed as shown in Fig. 7(d), and an anisotropic $\varepsilon_{zz}$ strain distribution can be observed.

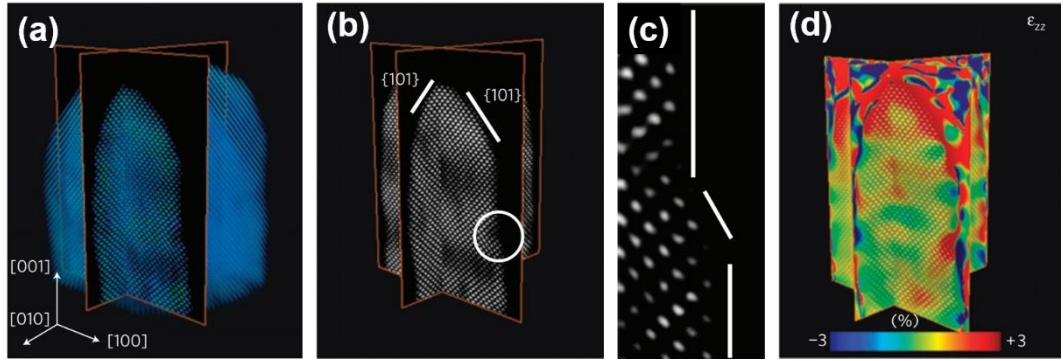

**Fig. 7. Atomic-resolution 3D reconstruction of an Au nanorod.** (a)The volume rendering image of the reconstructed Au nanorod. (b) Two orthogonal slices extracted from (a). (c)The region of a surface step circled in (b). (d) The corresponding 3D $\varepsilon_{zz}$ strain measurement of the two orthogonal slices in (a).[48]

Atomic resolution 3D reconstruction based on discrete tomography requires multiple high resolution projection images acquired along specific zone axes of the sample, which would not be feasible for samples such as nanodecahedra particles, because only one suitable axis can be available in such structure. To overcome this problem, Goris et al. developed a novel reconstruction algorithm based on modified SIRT reconstruction, where each atom in the Au nanodecahedron is modeled by a 3D

Gaussian function and the 3D Gaussian model is further introduced into the reconstruction as prior knowledge.[49] The most important advantage of this method is that all atom coordinates can be obtained and put out directly to calculate the 3D displacement map. Figure 8(a) shows the 2D analysis of strain distribution from a STEM-HAADF [110] projection image, where no systematic variation of the lattice parameter is observed from the resulting $\varepsilon_{xx}$ (Fig. 8(b)). However, from the 3D reconstruction results (Figs. 8(c) and 8(e)), a systematic outward expansion of the lattice occurs both along the $x$ and $z$ direction, as shown in Figs. 8(d) and 8(f). The difference between the 2D and 3D analysis of strain distribution along the $x$ direction is attributed to the averaged information obtained from the slices at different positions owning different $\varepsilon_{xx}$ strain distributions. The displacements along the z direction can only be extracted through a 3D approach. This comparison highlights the limitation of 2D projection images and emphasizes the importance of 3D reconstruction in characterizing nanoparticles with complex structural strain.

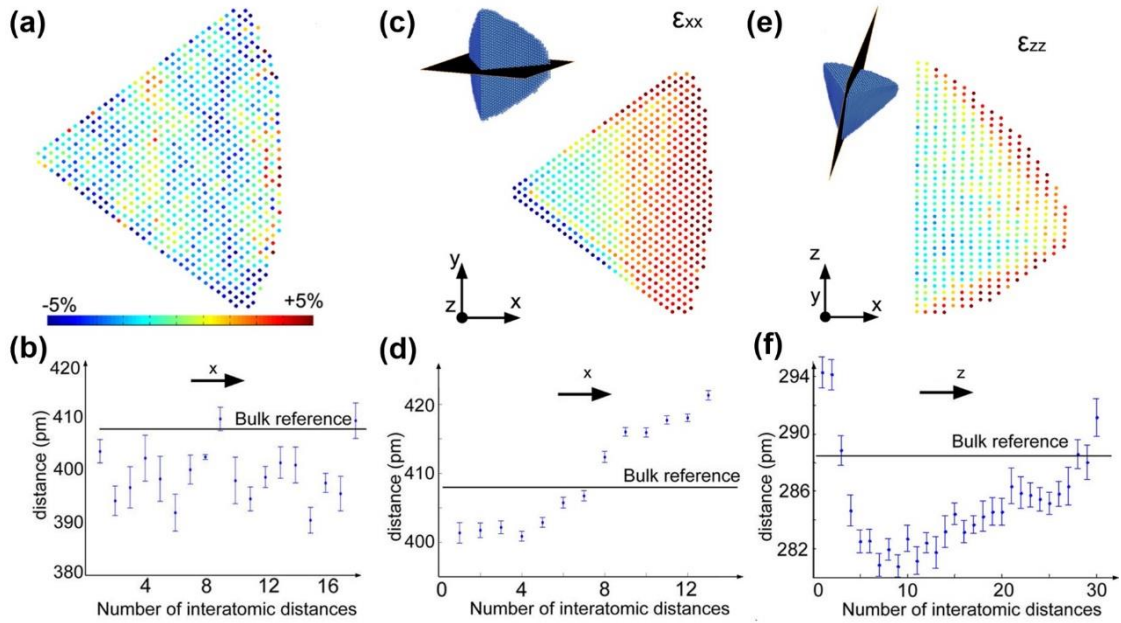

**Fig. 8. 2D and 3D analysis of strain distribution of an Au nanodecahedron.** The strain distribution is visualized using a color code and scaled between ± 5%. (a) The $\varepsilon_{xx}$ strain map obtained from a STEM-HAADF projection image. (b) The corresponding lattice displacement along the $x$ direction. (c) Slice through the $\varepsilon_{xx}$ volume from 3D

reconstruction. (d) The lattice displacement along the *x* direction corresponding to (c). (e) Slice through the $\varepsilon_{zz}$ volume from 3D reconstruction. (d) The corresponding lattice displacement along the *z* direction.[49]

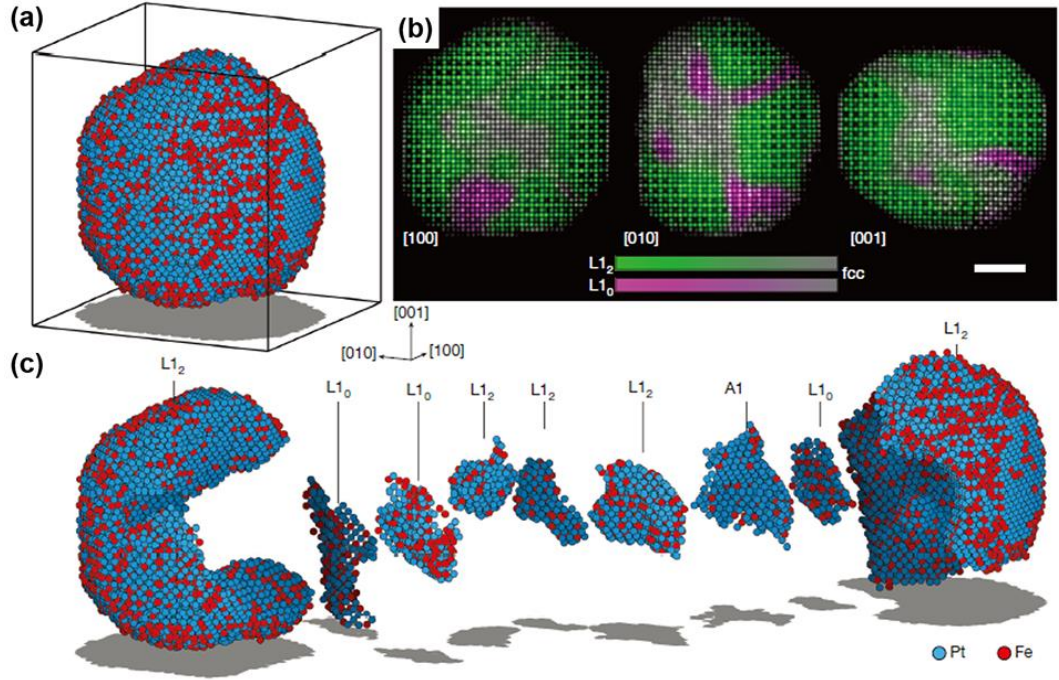

**Fig. 9. 3D reconstruction of an FePt nanoparticle including 3D atomic positions, elemental identification and detailed grain structure.** (a) The 3D positions of individual atoms of Fe and Pt. (b) Multislice images through the reconstructed 3D atomic model along the [100], [010] and [001] directions. Color bars indicate the degree of ordering, varying between pure $L1_2/L1_0$ and chemically disordered fcc. Scale bar, 2 nm. (c) The nanoparticle consists of two large $L1_2$ grains, three small $L1_2$ grains, three small $L1_0$ grains and a Pt-rich A1 grain.[50]

STEM-HAADF imaging provides electron tomography with a power of analyzing compositional/chemical information in three dimensions using its Z-contrast nature. Yang et at. reconstructed an FePt nanoparticle based on a series of STEM-HADDF images and proposed an atom tracing and classification method to determine the coordinates of all individual Fe and Pt atoms.[50] Figure 9(a) shows the spatial distribution of the Fe and Pt atoms in a FePt nanoparticle, identified by analyzing their local intensity distribution. Figure 9(b) shows the multislice STEM-HAADF simulation images based on the 3D atomic model along the [100], [010] and [001] directions, which exhibits several deceptive 'L1$_0$ grain' signatures from the overlapping of the two large L1$_2$ grains (labeled by magenta in Fig. 9(b)). More details from the 3D reconstruction indicate that the FePt nanoparticle has a complex structure with several short-range order parameter phases: two large L1$_2$ FePt$_3$ grains (chemically ordered face-centered cubic (fcc) phase) with interlocking concave shapes, three L1$_0$ FePt grains (ordered face-centred tetragonal phase), three small L1$_2$ FePt$_3$ grains and a Pt-rich A1 grain (chemically disordered fcc structure phase), as shown in Fig. 9(c). This example clearly demonstrates the unique power of atomic electron tomography in unveiling the high complexity of 3D chemical order/disorder in realistic samples. More importantly, the measured atomic coordinates with high precision by this method can be used as direct input for first-principles calculations for the understanding of material properties at the single-atom level.

In summary, 3D reconstruction techniques based on STEM-HAADF imaging are promising in revealing the 3D structural information, such as porous structure, surface facets, 3D lattice displacement and atomic coordinates, with a high spatial resolution. It is worth noting that the development of new algorithms such as artificial neural networks can further improve the efficiency of reconstruction by using very limited number of projection images and simultaneously guarantee the quality of the reconstructed objects.[51] The developments towards highly reliable 3D atomic-scale reconstruction will of no doubt advance our understanding of the structure-activity relationships in heterogeneous catalysts.[52, 53]

**Part 3. *In-situ* characterization of heterogeneous catalysts**

It is well known that structure of catalysts does change during catalytic reactions. Static structural information obtained under high vacuum and room temperature conditions in conventional (S)TEM may not represent the truly active structure for catalysis. Recording the structural and chemical evolution of catalytic materials under realistic reaction conditions is, therefore, of fundamental importance for unveiling the underlying mechanism and designing of novel catalysts with better performance. To bridge this gap, in-situ (S)TEM techniques have been developed to identify the intermediate structures and to capture structural evolution with atomic resolution under gas and heating conditions that mimic those in real catalytic reactions.

Two different approaches for in-situ (S)TEM experiments have been developed in order to accommodate the gas environment around the specimen into a TEM column that requires high-vacuum. The first approach involves major modification to the microscope column and uses differential pumping system to change a conventional TEM into an environmental TEM (EFEM),[54] as shown in Fig. 10(a). The designing principle of this open-type configuration limits the pressure in ETEM to much lower than the atmospheric pressure, typically not higher than 20 mbar.[54] Although the ultra-dilute reactive gases would cause the observed results deviating from those under real reaction conditions, the reduced reaction rate under the ultralow gaseous pressure provides feasibility to record the dynamical process during catalytic reactions with a relatively low temporal resolution achievable on most ETEM. In addition, the diluted gas helps to minimize the undesired scattering to the incoming electron beam by the gas molecule, and atomic resolution can still be obtained with ETEM imaging.[55, 56] A major drawback of the ETEM approach is that the addition of post-specimen differential pumping apertures blocks the high-angle scattering electrons, which makes it incompatible with STEM-HAADF imaging.[57]

The second approach uses specially designed microelectromechanical systems (MEMS)-based cells or nanoreactors, where a thin layer of gas is sealed between two electron-transparent windows, as shown in Figs. 10(b)-10(d).[55] These in-situ functional holders have become very popular over the past few years. They are flexible and can fit into almost any commercial (S)TEM without modification of the column. The key component of this technique is the electron-transparent window materials, which must have a low electron-scattering nature in order not to cause significant degradation of the spatial resolution due to multiple electron scattering and high fracture strength is also needed to withstand the inner and outer pressure difference. The most commonly used window materials are amorphous $SiN_x$ [58] and carbon film.[59]

With such in-situ holders, catalytic reaction under atmospheric pressure and high heating temperature (~1000°C) can be studied *in-situ*,[60, 61] although possible radiation damage from high-energy electron beams to the window films should be considered in practical applications.[62]

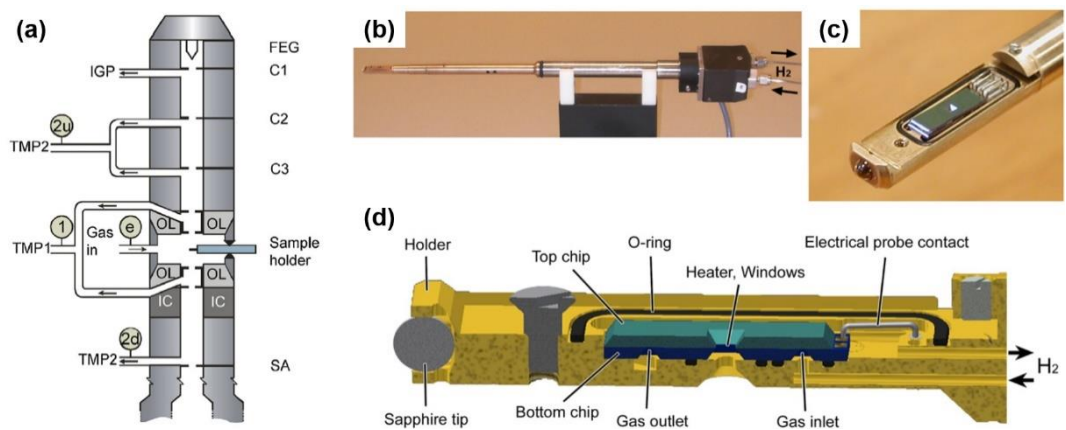

**Fig. 10. Two different approaches for in-situ experiments using either ETEM or MEMS-based functional holders.** (a) Schematic of a differential pumping system in ETEM.[54] (b-d) Configuration of nanoreactors in the in-situ sample holder.[55]

For heterogeneous catalysis with gas reactants, the reaction efficiency largely depends on the surface morphology of the nanocrystals while the reaction gases also affect the surface structure of the catalysts. Therefore, real-time TEM observation of the surface structure evolution at the atomic scale under reaction environments is of particular importance. Zhang et al. investigated the surface amorphization of anatase $TiO_2$ exposed to $H_2O$ vapor of 1 Torr at 150 °C at the atomic scale, simulating the conditions of vapor-phase water splitting.[63] The *in-situ* experiment was conducted in a differentially pumped ETEM and a series of HRTEM images were recorded as shown in Fig. 11. The sample without exposing to $H_2O$ vapor exhibited clear {101} surface (in Fig. 11(a)), while a disordered layer formed and thickened along the {101} facet with electron beam irradiation under 1 Torr $H_2O$ vapor, as highlighted in Figs. 11(b)-11(e). In comparison, the sample only exposed to water vapor for 40 hours at 150 °C was also imaged (Fig. 11(f)) and clean crystalline {101} surface was observed. Similar amorphization of the $TiO_2$ surface was observed when exposed to $H_2O$ vapor and ultraviolet light, and it is proven that amorphous titania hydroxide layers on $TiO_2$ surface facilitate $H_2O$ molecule adsorption and promote the water splitting.

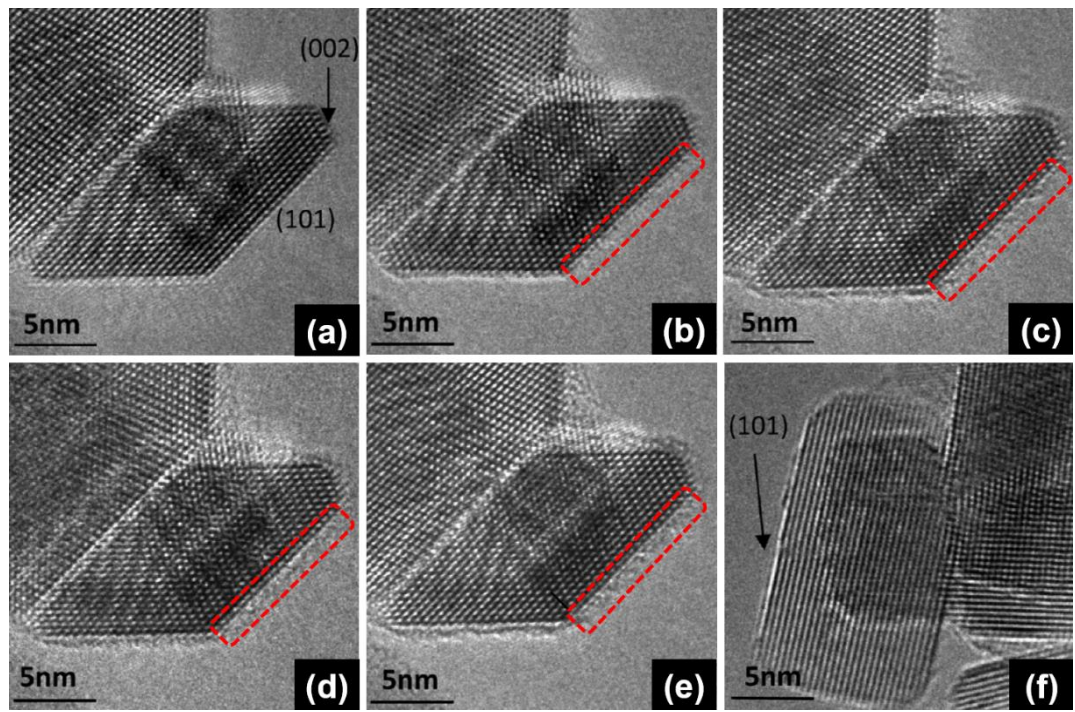

**Fig. 11. *In-situ* HRTEM imaging of anatase nanocrystals at 150 °C.** (a) no $H_2O$ vapor; (b-e) exposed to 1 Torr $H_2O$ vapor after 1 h (b), 7 h (c), 20 h (d), and 40 h (e); (f) fresh areas only exposed to $H_2O$ vapor after 40 h without electron irradiation.[63]

Surface oxidation and reduction often occur during catalytic reactions. Xin et al. used ETEM to study the structural evolution of Pt-Co nanoparticles during *in-situ* annealing under $O_2$ or $H_2$ environment.[64] As shown in Figs. 12(a)-12(c), CoO-phase island forms and grows gradually onto the surface of the Pt-Co nanoparticles under oxygen environment. Figures 12(d)-12(f) present the corresponding lattice spacings measured from the atomic resolution TEM images. The lattice expansion is attributed to the infiltration of oxygen into the Pt-Co lattice, and the CoO phase segregates onto the surface after 12s of reaction. The CoO segregation was found to be reversible during the *in-situ* reduction under $H_2$ atmosphere. The reduction behavior of the oxidized Pt−Co nanoparticles was monitored using STEM low-angle annular dark field (LAADF) imaging as shown in Fig. 12(g). During the early reduction reaction stages with $H_2$ (Fig. 12(g), 241-503 s), the CoO on surface was reduced and formed new Co metallic nanoparticles. Subsequently, Co migrated back and reincorporated into the platinum-rich particles again, as can be seen from the disappearance of the Co nanoparticles and the associated shape change of the larger Pt-Co particles. The segregation of Co onto the surface during oxidation and reincorporation into the alloy particles during reduction, as revealed by the in-situ (S)TEM studies, help to track the accurate intermediate steps of the reaction pathway.

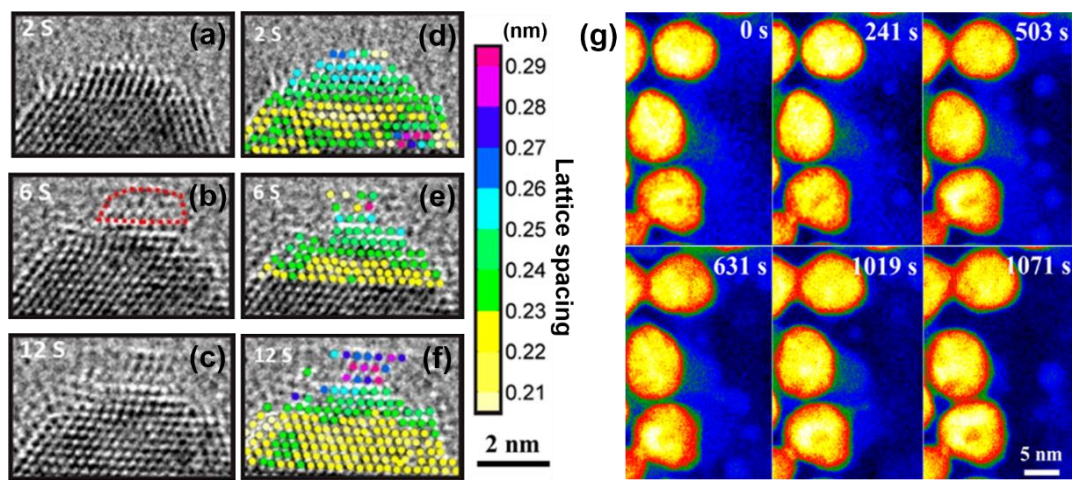

**Fig. 12. *In-situ* characterization of $Pt_{0.5}Co_{0.5}$ nanocrystals during oxidation and reduction reactions.** HRTEM images showing the CoO-island forming behavior in a $Pt_{0.5}Co_{0.5}$ nanocrystal under 0.1 mbar $O_2$ and 250 °C at 2 s (a), 6 s (b) and 12 s (c), and the corresponding lattice spacing as measured from the HRTEM images (d-f). (g) STEM-LAADF images for the *in-situ* reduction of oxidized Pt-Co nanoparticles under $H_2$ at 400 °C (CoO in blue and the metallic core in yellow).[64]

The limited pressure around the specimens achievable in an ETEM limits the number of catalytic reactions that can be studied *in-situ* and causes a knowledge gap between the real working behaviors of catalysts under atmospheric conditions and the results obtained using ETEM. Recent technical advances in nanofabrication of nanoreactors have allowed for atomic resolution *in-situ* (S)TEM imaging under atmospheric pressure using the state-of-the-art *in-situ* gas holders, offering new opportunities to probe the catalyst structure under more realistic reaction conditions.

Using such MEMS-based *in-situ* gas/heating holders, Vendelbo at el. observed an oscillatory behavior of Pt nanoparticles when catalyzing CO oxidation under one atmosphere pressure and heating to 659 K.[65] By simultaneously monitoring the CO pressure (using mass spectroscopy, shown in Fig. 13(f)), the reaction power and the structure of the Pt nanoparticles, it was found that as the CO conversion rate increased rapidly the Pt nanoparticles transformed from a more spherical shape towards a more facetted shape. This shape transformation reversed when the CO conversion rate decreased, and the more spherical shape retained until the CO conversion increased again, as shown in Figs. 13(a)-13(e). More detailed HRTEM images in Fig. 13(g) show that, in the more spherical state, more open (110) planes and step sites were observed, while a reduction of higher index terminations and steps and an increase of the close-packed (111) planes were observed in the more facetted particles. Combining the simultaneous *in-situ* mass spectroscopy and HRTEM studies, it was concluded that the oscillatory evolution of surface morphology is synchronous with the change in the CO conversion rate, and higher conversion is associated with a transformation towards more close-packed (111) facets.

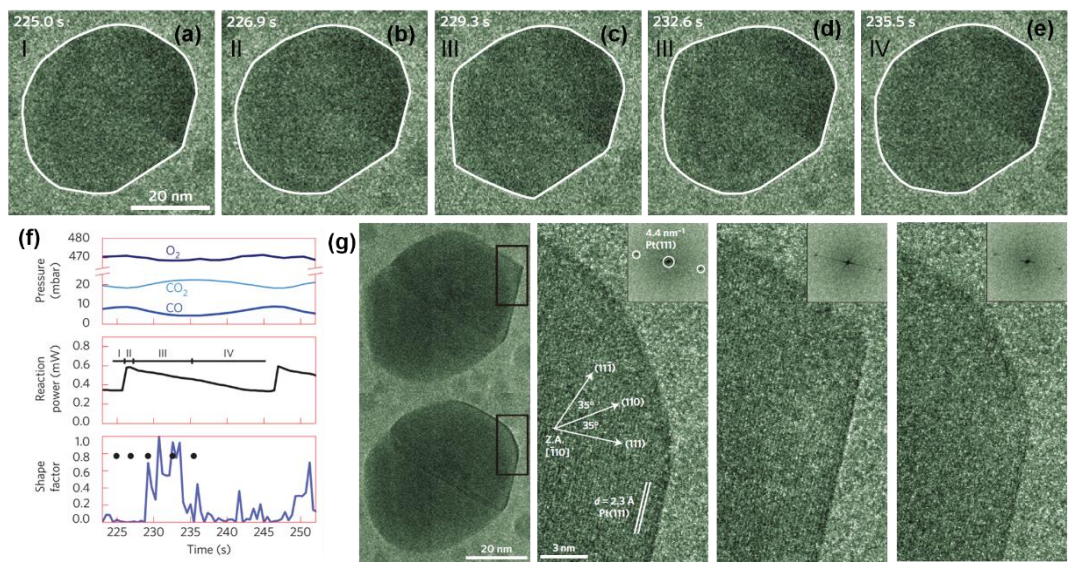

**Fig. 13. *in-situ* HTREM imaging of oscillatory behavior for Pt nanoparticles during CO oxidation reaction.** (a-e) A series of *in-situ* HRTEM images of the morphology evolution of a Pt nanocrystal showing an oscillatory behavior. (f) Real-time mass spectra for probing $O_2$, $CO_2$ and CO pressures, reaction power, and shape factor. (g) Atomic-scale imaging of a single Pt particle with different shapes: the more spherical shape and the more facetted shape.[65]

Surface faceting behavior was also observed in PdCu nanocrystals under $H_2$ conditions. Jiang et al. used *in-situ* gas/heating holder to reveal the morphological evolution of PdCu nanocrystals when annealed under the $H_2$ pressure of one atmosphere at 600K.[66] As the annealing proceeded, the spherical nanocrystal rotated and transformed towards a morphology containing more facets with smaller curvature. Eventually, four flat (001) facets emerged and the nanocrystal transformed into a truncated cube with (001) and (011) facets, shown in Figs. 14(a)-14(d). The corresponding FFT patterns in Figs. 14(e)-14(h) show the evolution of crystalline facets during the annealing process. In contrast, the nanocrystals maintained their original spherical shape when annealed under lower $H_2$ pressure (0.016 bar). By combining these *in-situ* experimental results with the first principle calculations, it was reported that the distinct faceting transformation can be attributed to a new range of the surface energy as $\gamma_{H-001}$ (0.90 J m$^{-2}$) <$\gamma_{H-010}$ (0.99 J m$^{-2}$) <$\gamma_{H-111}$ (1.06 J m$^{-2}$) at 1 bar $H_2$ pressure.

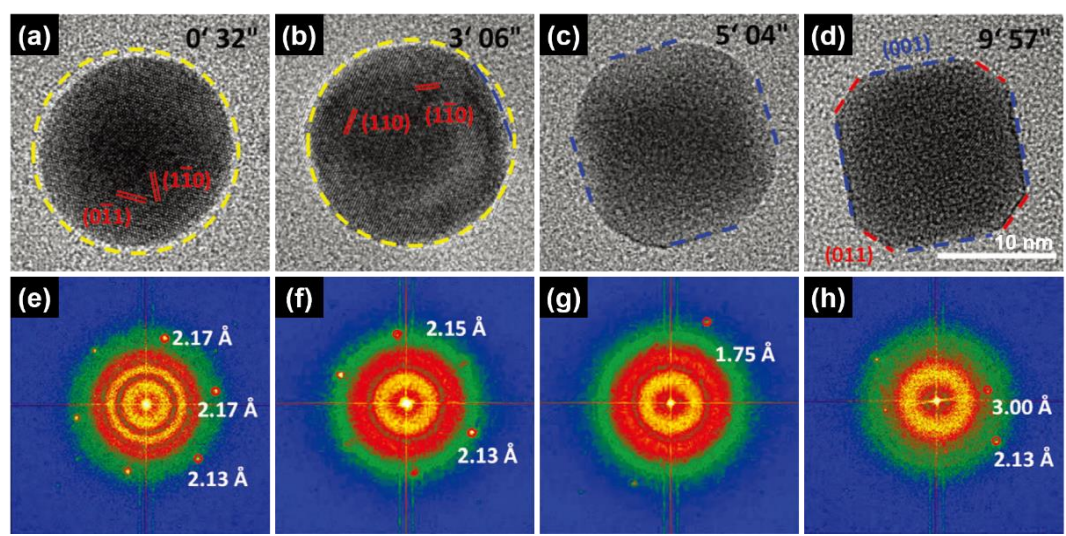

**Fig. 14. A series of *in-situ* HRTEM imaging and FFT analysis of a PdCd nanocrystal during annealing under $H_2$ of 1 atmosphere pressure at 600 K.** (a-d) HRTEM images of the surface faceting of the PdCd nanocrystal and (e-h) the corresponding FFT patterns.[66]

The incorporation of nanoreactors into TEM not only drastically increases the gas pressure achievable for *in-situ* experiments, but also makes it possible to perform *in-situ* STEM-HAADF imaging, which is not feasible in ETEM. For alloy nanocatalysts, the distribution of elements and their diffusion behaviors during thermal annealing are crucial to reveal the structure-property relationship and for tailoring their catalytic performance.[67, 68] The implement of *in-situ* STEM-HAADF imaging provides an excellent way to analyze the compositional evolution of alloy nanoparticles and elucidate the underlying mechanism responsible for these changes. Using a MEMS-type *in-situ* heating holder, Chi at el. studied the temperature-dependent Pt-segregation in $Pt_3Co$ nanoparticles during thermal annealing.[69] Owing to the Z-contrast in the STEM-HAADF images, the segregation of Pt can be easily observed from the increase of image intensity. As shown in Fig. 15, when annealed at 350 °C an obvious increase of image intensity can be observed in the outmost surface atomic layer of $Pt_3Co$ particles (Fig. 15(e)), which disappeared at higher annealing temperature of 550 °C. Assisted with multi-slice imaging simulation and EDS mapping, the formation of a Pt-rich layer on the surface of $Pt_3Co$ nanoparticles was confirmed.

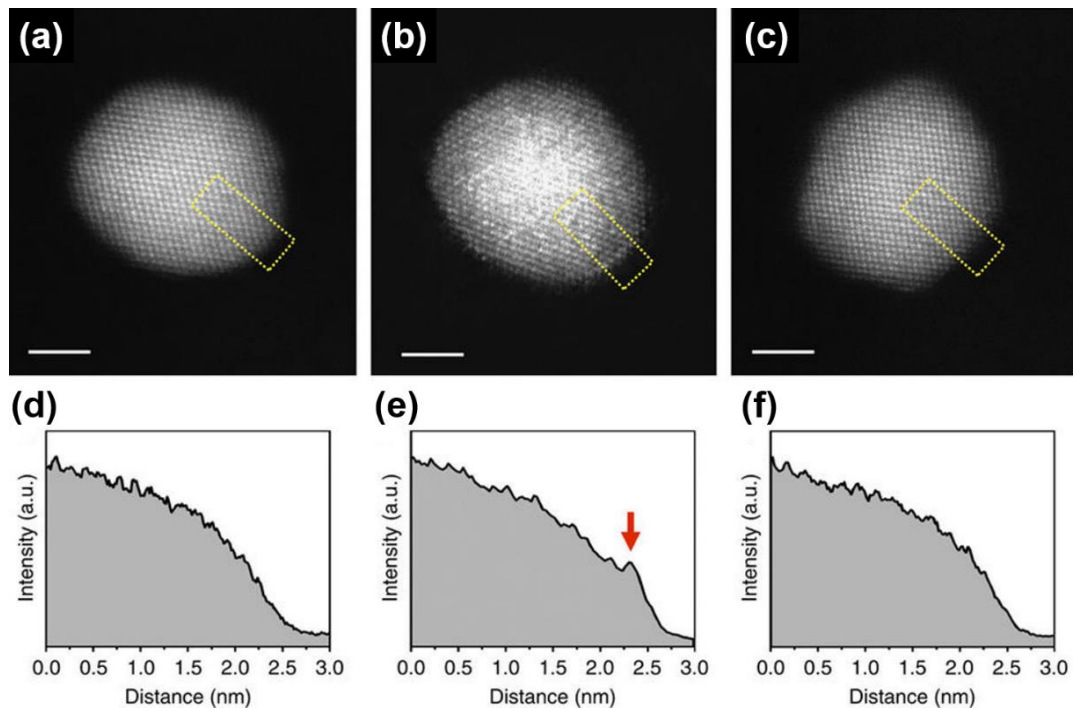

**Fig. 15. STEM-HAADF imaging of a single $Pt_3Co$ particle during *in-situ* annealing.** (a-c) STEM-HAADF images at room temperature (a), 350 °C (b) and 550 °C (c). (d-f) The corresponding intensity profiles of images a-c along the yellow dotted boxes.[69]

Even though the MEMS-based *in-situ* holders have only been developed for a few years, they have allowed for atomic resolution imaging under both TEM and STEM modes under atmosphere pressure and high heating temperature (>1000°C). Using an *in-situ* gas/heating holder, Dai at el. revealed how different elements segregated onto different facets on Pt$_3$Co nanocrystals when annealed at 350 °C under pure oxygen environment of 760 Torr.[70] Figure 16(a) shows a STEM-HAADF image of a particle after annealing with the corresponding intensity line profiles in Fig. 16(b). It is clear that Co atoms diffuse from the bulk onto the surface and form a Co-rich layer on the {111} surfaces during annealing. The STEM-BF image in Fig. 16(c) demonstrates the formation of CoO layers along {111} surface of Pt$_3$Co, as further confirmed by lattice spacing measurement and EELS analysis. In contrast, segregation and oxidation of Co was not observed on the adjacent {100} facets of Pt$_3$Co, leaving clean {100} surfaces. Sequential STEM images (Figs. 16(d)-16(g)) further revealed a distinct diffusion behavior of Pt atoms exclusively along the clean Pt$_3$Co {100} facets, which is attributed to the formation of volatile Pt-oxygen species in an O$_2$ atmosphere whereas the CoO layers on {111} facets block the underlying Pt atoms from exposure to the oxygen environment. These *in-situ* results provide direct atomic-scale insights into elemental diffusion and phase transformation under realistic reactions conditions involving gas environment and heating. The readers are encouraged to refer to recent reviews on topics such as *in-situ* (S)TEM under atmospheric pressures or gas-involved *in-situ* TEM.[71, 72]

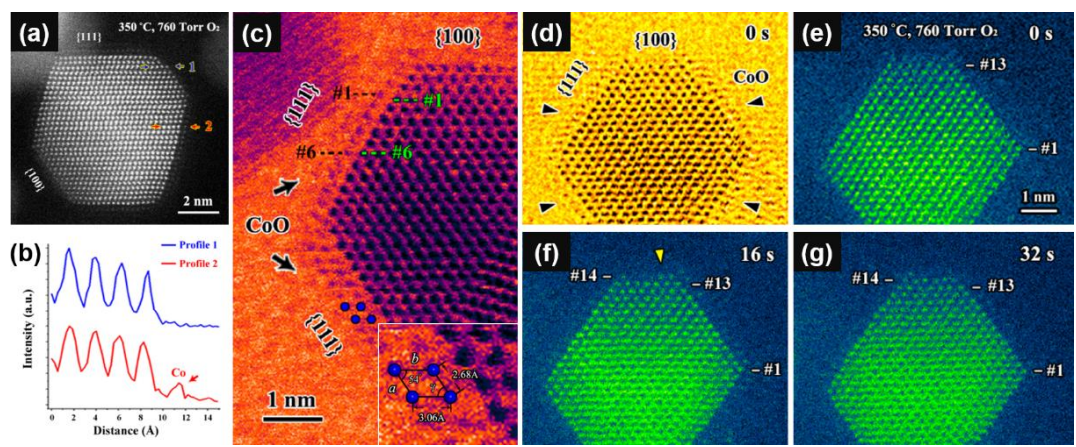

**Fig. 16. *In-situ* STEM imaging of Pt$_3$Co nanocrystals under oxygen environment and elevated temperature.** (a) STEM-HAADF image of a Pt$_3$Co nanocrystal after annealing and (b) the corresponding intensity line profiles. (c) False-colored STEM-BF image showing the formation of CoO layers on the Pt$_3$Co {111} surfaces and the corresponding schematic of Co atom arrangement along the CoO <110> zone axis (inset). (d-g) Sequential STEM images showing the diffusion and formation of a Pt atomic layer on the {100} surface of an oxidized Pt$_3$Co particle.[70]

**Outlook**

There is no doubt that atomic resolution analytical (S)TEM techniques have played an important role in the study of heterogeneous catalysts. The atomic scale imaging and spectroscopy information obtained from the state-of-the-art (S)TEMs provide a unique way to probe the local structure of the catalysts, complementing the averaged structural information obtained from optical and X-ray techniques. Meanwhile, it should always be kept in mind that the high-energy electron beam may well change the intrinsic structure of the catalyst sample due to irradiation damage. This irradiation effect should be very carefully taken into account when studying the intrinsic structure of catalysts, especially under gas or liquid environments. Even though the recent advances in low-voltage low-electron-dose (S)TEM and high sensitivity detectors has helped to minimize the beam damage effect, irradiation damage is still a major challenge for the study of beam-sensitive materials such as 2D materials, metal-organic frameworks (MOFs) and zeolites. The extra electron dose spent during searching for crystal zone axis, setting up the right imaging conditions and adjusting the accurate defocus value also exacerbate the damage. The newly developed direct electron detection cameras are expected to overcome this challenge by drastically reducing the electron dose required for high resolution imaging.[73, 74] Further combined with sophisticated algorithms and imaging techniques, such as compressive sensing, inpainting and ptychography, the development towards ultra-low-electron-dose imaging should enable direct imaging of extremely beam sensitive materials at the atomic level. Furthermore, cryo-TEM (cryogenic transmission electron microscopy) is another promising tool to probe electron-beam sensitive specimens. Cryo-TEM has been extensively used in the study of biological samples, and has also found important applications in material science.[75, 76] Under low temperature (approximately 100 K) and low-dose (<100 e Å$^{-2}$) conditions, the irradiation damage can be minimized and the intrinsic structure can be better retained for prolonged TEM observation.

As discussed in the second part of this review article, STEM-HAADF tomography can now achieve atomic resolution in 3D owing to advances both in instrumentation and reconstruction algorithms. However, using the Z-contrast of STEM-HAADF imaging to identify and separate the spatial distribution of different elements in tomography reconstruction is still challenging, especially when the elements have similar atomic numbers. Combining atomic-resolution spectroscopy (EDXS or EELS) and tomography offers a possible way to reconstruct the 3D structure with detailed chemical information and atomic resolution. For this purpose, new techniques and data processing algorithms must be developed to drastically cut off the electron dose required for tomography with spectroscopy signals in order for the catalyst particles to survive during the experiments.

The recent development of *in-situ* techniques has allowed for direct imaging of the dynamically structural and compositional evolution of catalysts at the atomic scale under more realistic reaction conditions involving gas pressure and elevated temperatures. However, temporal resolution is still a bottleneck for *in-situ* (S)TEM for the study of dynamical events, whereas catalytic reactions usually feature high-frequency dynamics such as nucleation, atomic diffusion and surface reconstruction. Thus, combining ultra-fast electron microscopy[77, 78] with *in-situ* techniques and fast cameras may provide a promising solution to improve the temporal resolution of (S)TEM for the study of structural dynamics during catalytic reactions. However, it should be kept in mind that the structural changes under the excitation of ultra-fast pulse laser may deviate from the intrinsic structural evolution occurring under realistic catalytic reactions, and the intense electron pulse might also cause damage to the catalyst particles (if working under dynamic TEM (DTEM) mode[79, 80]). Special cares will need to be taken when probing the fast dynamics of catalysts using ultra-fast electron microscopy.

**Acknowledgements:**

This research was supported by the Natural Science Foundation of China (51622211) and the CAS Pioneer Hundred Talents Program.